\documentclass[onecolumn,showpacs,superscriptaddress,amsmath,amssymb,nofootinbib]{revtex4}

\usepackage{epsfig,float,amsmath,amstext,amssymb,amsfonts,bm}

\usepackage{dcolumn}
\usepackage{bm}

\newcommand{\boldl}{{\bf L}}
\newcommand{\boldu}{{\bf U}}
\newcommand{\tr}{{\rm tr}\,}

\begin{document}

\title{Lattice simulation of a center symmetric
three-dimensional effective theory for SU(2) Yang-Mills}

\author{Dominik Smith\footnote{I thank A.~Dumitru and S.~Schramm
for their collaboration and helpful discussions, and R.~Pisarski and
Zs.~Schram for useful comments and suggestions. I gratefully acknowledge
a fellowship by FIAS and support by the Helmholtz foundation.}}

\affiliation{
Institut f\"ur Theoretische Physik,
Johann Wolfgang Goethe-Universit\"at,
Max-von-Laue-Str.\ 1, 60438  Frankfurt am Main, Germany
}

\date{\today}
\begin{abstract}
We perform simulations of an effective theory of SU(2) Wilson lines in
three dimensions. Our action includes a kinetic term, the one-loop
perturbative potential for the Wilson line, a non-perturbative ``fuzzy-bag''
contribution and spatial gauge fields. We determine the phase
diagram of the theory and confirm that, at moderately weak coupling,
the non-perturbative term leads to eigenvalue repulsion in a finite region above the deconfining phase transition.
\end{abstract}
\maketitle

\section{Introduction}
Recent results from the Relativistic Heavy Ion Collider (RHIC)
demonstrate qualitatively new behavior for heavy-ion collisions at
high energies.  RHIC appears to probe a regime where quarks and gluons
are no longer bound into hadrons but form a new state of matter, a
Quark-Gluon Plasma (QGP). The experimental results suggest that the
transition does not occur directly from a confined phase to an ideal
QGP but that it goes through a phase where the deconfined phase is
non-perturbative~\cite{Gyulassi:2005}. In fact, ``partial''
deconfinement is expected from the Gross-Witten matrix model for
SU($N$) Wilson lines at large $N$~\cite{GWmatrix} and from
quasi-particle approaches~\cite{QP}. Here, we
investigate an effective theory of straight, thermal Wilson
lines~\cite{Pisarski:2006hz} which respects the global Z(N) symmetry
of Yang-Mills and might be able to describe the region below about
$T\approx 3T_d$, where resummed perturbation theory fails. We focus,
in particular, on the distribution of the {\it Eigenvalues} of the SU(2)
Wilson line at, and slightly above, the deconfining phase transition.

\section{Effective theory}
\label{eff_th}
We consider an effective theory in three dimensions which uses the
thermal Wilson line $\boldl(\bm{x})$ as the fundamental degree of
freedom in the electric sector. This is a matrix valued field which
exists in three dimensional space. The effective Lagrangian is composed of
a kinetic term, the one-loop potential, and a
non-perturbative term proportional to a ``fuzzy-bag'' constant $B_f$:
\begin{align}
 	\label{Leff_cl}
{\cal L}^{\rm eff} = \frac{1}{2}\tr G^2_{ij} + \frac{T^2}{g^2}
   \tr |\boldl^\dagger D_i \boldl|^2 - \frac{2}{\pi^2} T^4
  \sum\limits_{n\ge1} \frac{1}{n^4} |\tr \boldl^n|^2 + B_f T^2 |\tr\boldl|^2~.
	\notag	
 \end{align}
$G_{ij}$ is the non-Abelian magnetic field strength and $D_i$ the covariant
derivative. This theory does not involve elements of the algebra, i.e.\ $A_0$, and respects the Z(N) center symmetry of the SU(N) gauge group. It is therefore able to generate fluctuations
between different Z(N) vacua, just above the phase transition.
The Lagrangian is non-renormalizable in three dimensions and thus applies only at distances $\gg 1/T$.
 
\section{Lattice simulation}
We performed simulations of the effective theory for two colors, employing
three-dimensional cubic
lattices with periodic boundary conditions. We use Metropolis and
overrelaxation techniques \cite{Metr,Ora} to update the lattice configurations and to generate a thermal ensemble.
We currently neglect all but the $n=1$ term of the perturbative potential which 
can then be combined with the non-perturbative term. The lattice action then reads
\begin{equation}
S= \beta \sum_\Box ( 1-\frac{1}{2} \rm {ReTr} \boldu_\Box )
 - \frac{1}{2}\beta\sum\limits_{\langle ij\rangle}
              \tr ( \boldl_i \boldu_{ij}\boldl_j^\dagger \boldu_{ij}^\dagger + {\rm h.c.}) 
        - m^2 \sum\limits_i |\tr\boldl_i|^2~.\notag
\end{equation}
It contains the standard kinetic Wilson action for the magnetic fields, where
the sum runs over all "plaquettes", gauge-covariant kinetic nearest neighbour interaction
for the Wilson lines and a mass term. The index $i$ labels lattice sites
and $\langle ij\rangle$ labels links.

Initial results were obtained by neglecting the magnetic sector \cite{DS}. 
The theory is then essentially reduced to a sigma model with global $\mathrm{SU}_L(2)\times \mathrm{SU}_R(2)$ symmetry. When $m^2=0$, a
second order phase transition occurs at $\beta_c = 0.942(5)~$.
Here, the expectation value of the ``length'' of $\overline\boldl$,
$u = \sqrt{\tr\overline\boldl^\dagger \overline\boldl/2}$,
takes on a non-zero value~\cite{KSS}.
The effective screening mass, which equals the inverse spatial correlation length,
drops to zero here (up to finite lattice size effects) and remains zero above the phase transition point,
due to massless Goldstone modes.

Choosing a fixed value for $\beta$ and varying $m^2$ also
gives a second order phase transition. Here, the proper order parameter is the expectation
value of the Polyakov loop, which is the trace of the Wilson line.
Larger $\beta$ shifts the phase transition point to smaller $m^2$. To obtain non-zero
VEVs for the Polyakov loop for negative $m^2$, an infinitesimal background field must be added. These results
suggest that the phase boundary of the theory runs diagonally in the $\beta$-$m^2$ plane~\cite{DS}.

We measured the average and the difference of eigenvalues of the Wilson lines
\begin{eqnarray}
\rho_1(t,\bm{x}) &=& \frac{1}{2}\left|\lambda_1(t,\bm{x}) -
                             \lambda_2(t,\bm{x})\right|~~,~~
\rho_2(t,\bm{x}) = \frac{1}{2}\left|\lambda_1(t,\bm{x}) + 
                             \lambda_2(t,\bm{x})\right|~\notag
\end{eqnarray}
and determined their distribution in the thermal ensemble over a broad range of couplings.
We find, that for $m^2=0$ and small $\beta$, the distribution is dominated by the
group integration measure which leads to a
logarithmic divergence for $\rho_1 \to 0$. For large $m^2$ one approaches the perturbative vacuum,
where the distribution of the average $\rho_2$ is peaked around 1. In the 
confined phase at large $\beta$ the Wilson lines fluctuate about the non-trivial vacuum $\boldl_c=i\tau_3$, or
SU(2) rotations thereof (eigenvalues ``repel'' ). Fluctuations diminish with increasing $\beta$.

\begin{figure}
\includegraphics*[width=0.49\linewidth]{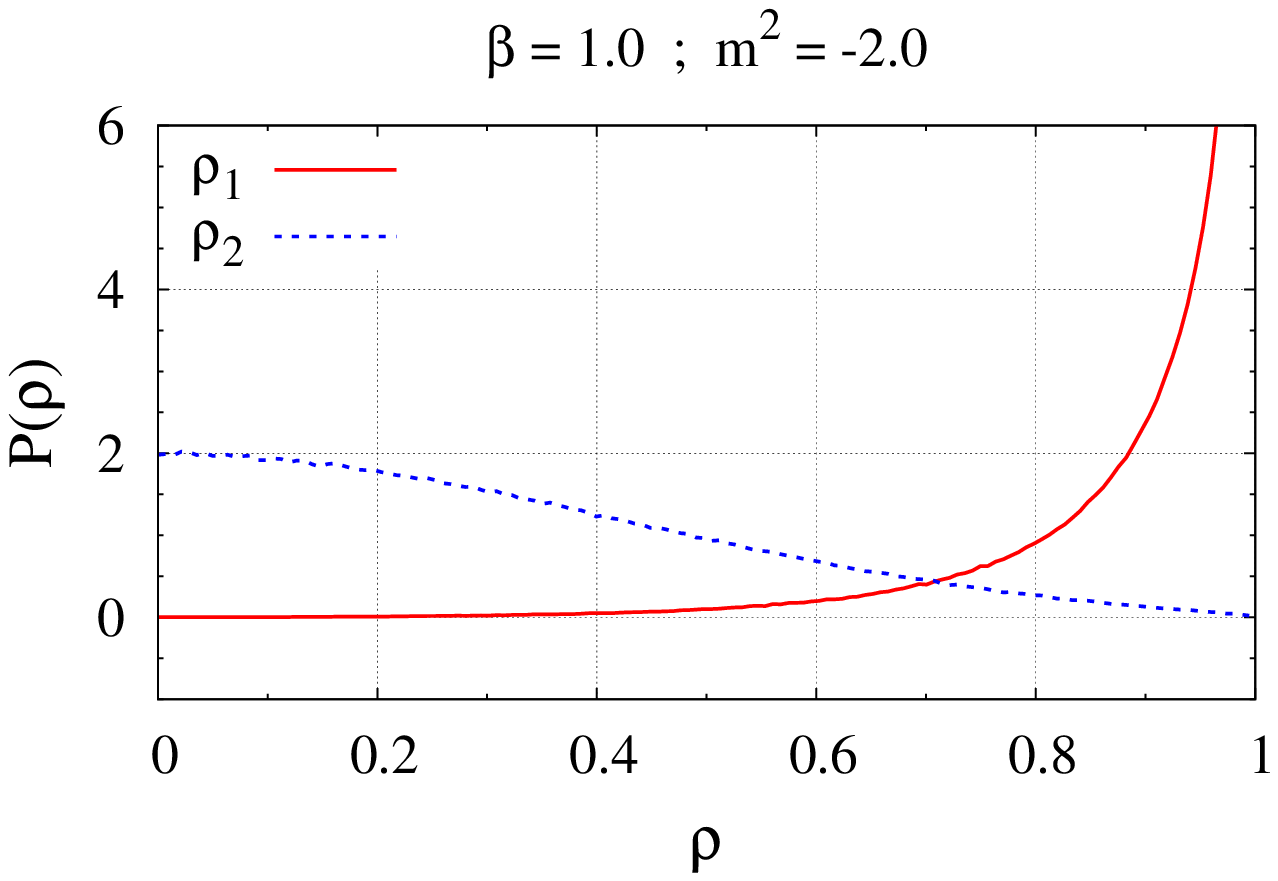}
\includegraphics*[width=0.49\linewidth]{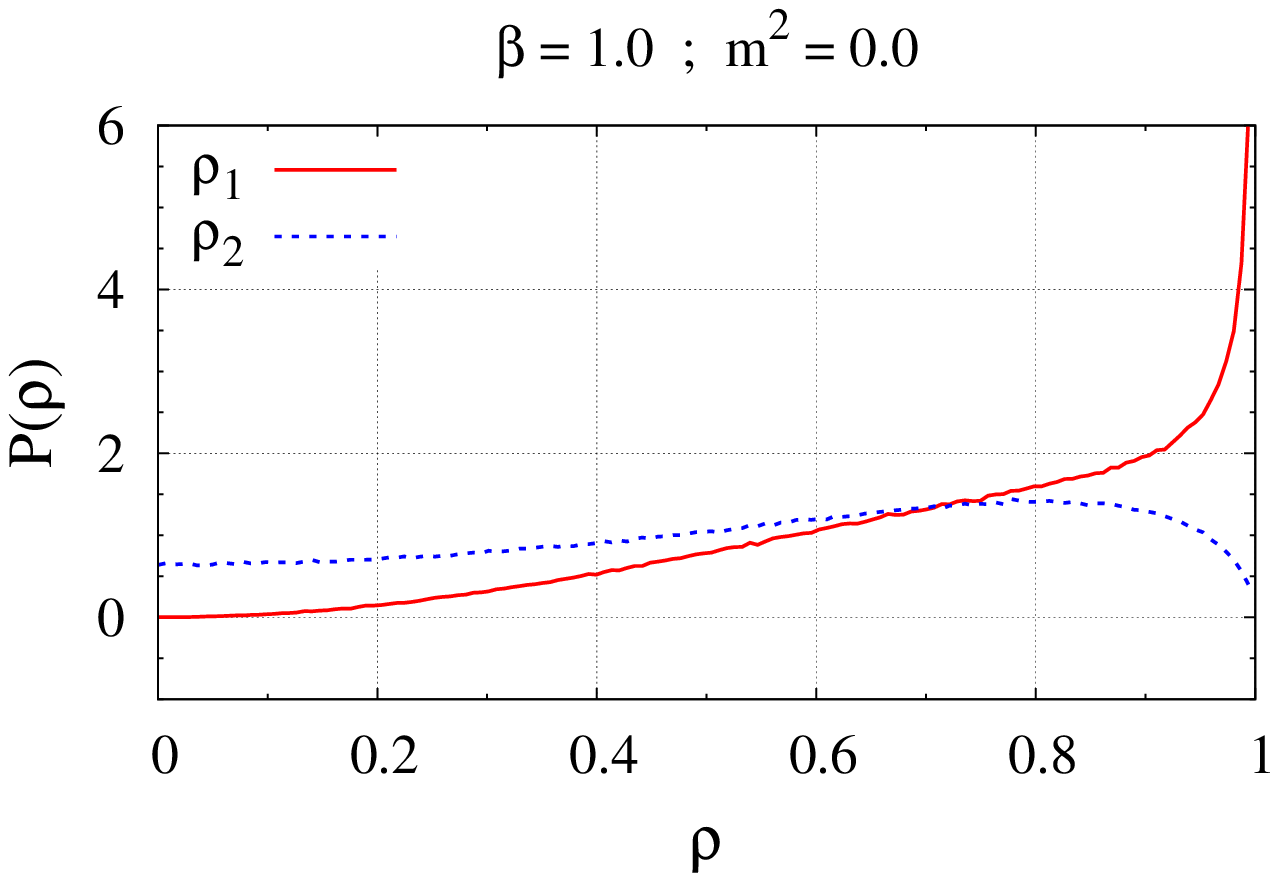}
\includegraphics*[width=0.49\linewidth]{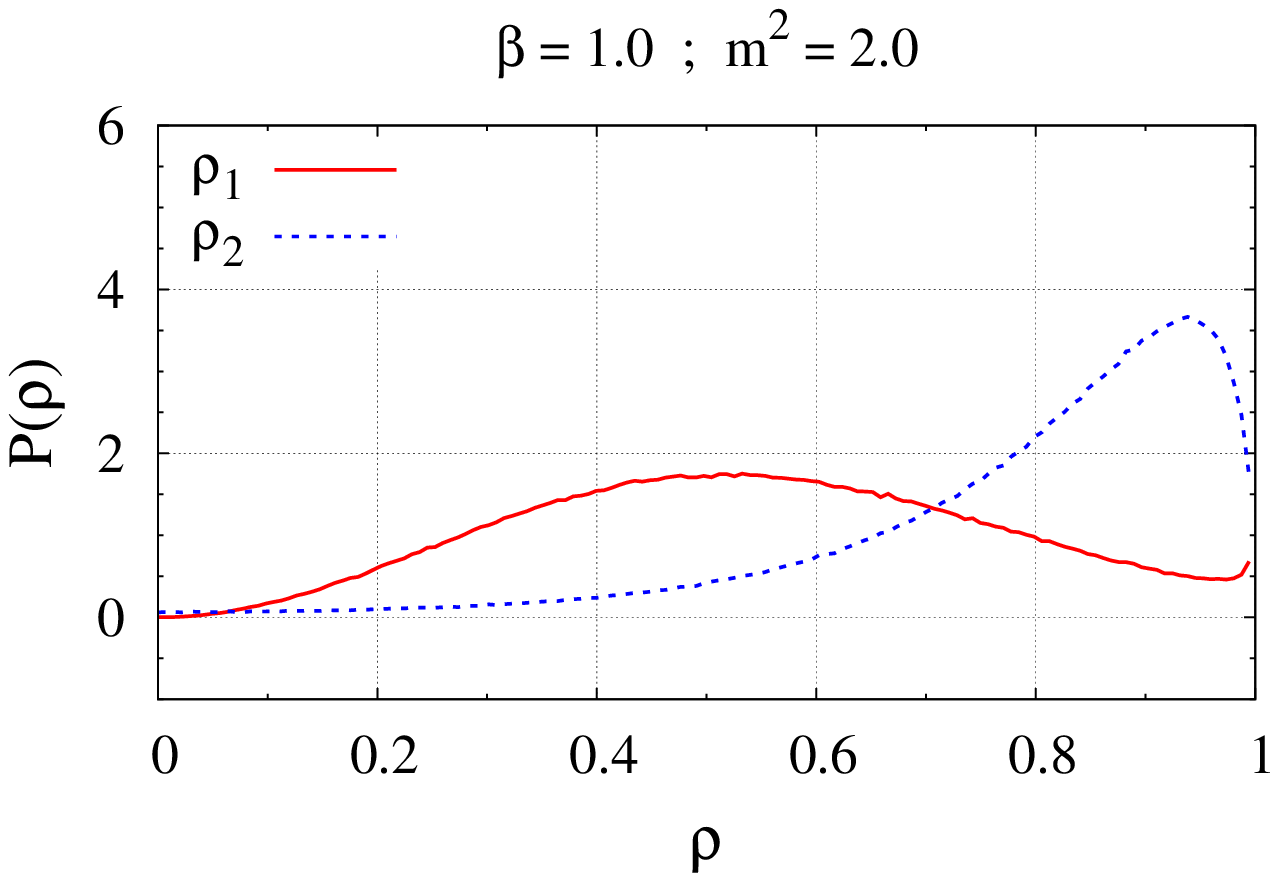}
\includegraphics*[width=0.49\linewidth]{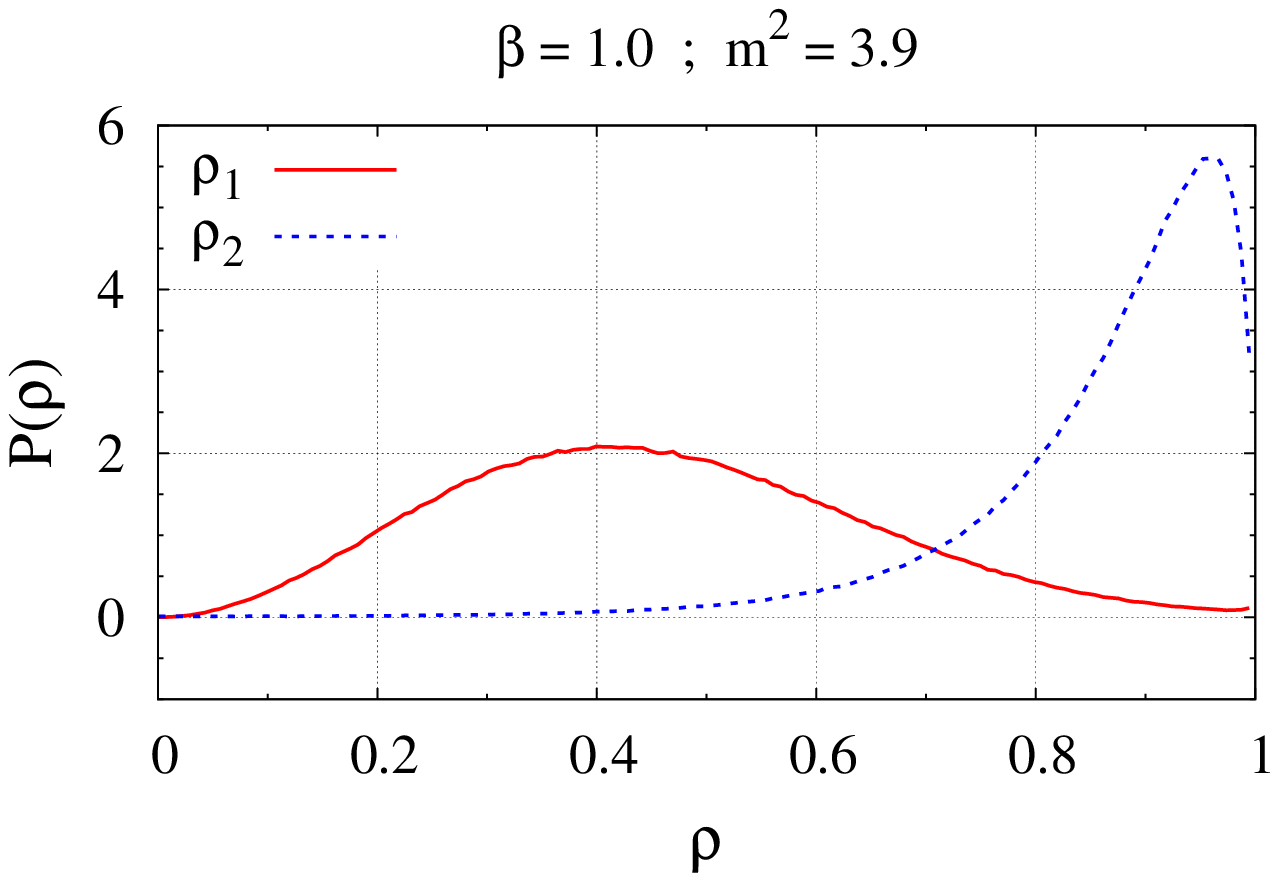}
\caption{Eigenvalue distributions measured on a $24^3$ lattice for different values of $m^2$ and
$\beta$. Magnetic fields are included. \label{fig:histo_fig}}
\end{figure}

Adding gauge fields now leads to a few qualitative changes in the observables.
Foremost, it is no longer necessary to use the 0(4)-length as an order parameter,
since the presence of gauge fields explicitly breaks the ${\rm SU}_L(2) \times {\rm SU}_R(2)$
symmetry of the kinetic term. The Polyakov loop is now a good order parameter, even for $m^2=0$.
All massless Goldstone modes disappear as well,
so that the inverse correlation length is only zero (again up to finite size effects)
exactly at the phase transition point.

Qualitative differences in the distribution of eigenvalues arise from
the fact that even at $m^2=0$ raising $\beta$ can now lead to
eigenvalue attraction. The qualitative dependence on $m^2$ for fixed
$\beta$ is similar to the spin-model case, although the transition from the
perturbative vacuum to the confined vacuum is smoother
across the point $m^2=0$ (as one would expect, since this is
reminiscent of the change of the behavior of the Polyakov loop).
Overall, changes in the electric
sector due to magnetic fields are qualitatively similar to the effect
of the external field which was formerly used to
find the phase boundary in the spin model at $m^2<0$.

To assess the influence of the dynamics of the Wilson lines on the
magnetic fields we have also measured the expectation value of the
Wilson action itself and of the spatial string tension. This was done by obtaining expectation values of rectangular
Wilson loops $W(I,J)$, where I and J are the dimensions of the loop in
lattice units. Such loops obey the area law $W(i,j) \propto \exp~(-\sigma A)$,
where $A=a^2IJ$ is the area of the loop and the string tension $\sigma$ plays the role of a decay constant. One normally uses Creutz ratios~\cite{Creutz} but our comparison of both methods suggests that in 3D, using the Wilson loops directly works equally well.

\begin{figure}
\includegraphics*[width=0.49\linewidth]{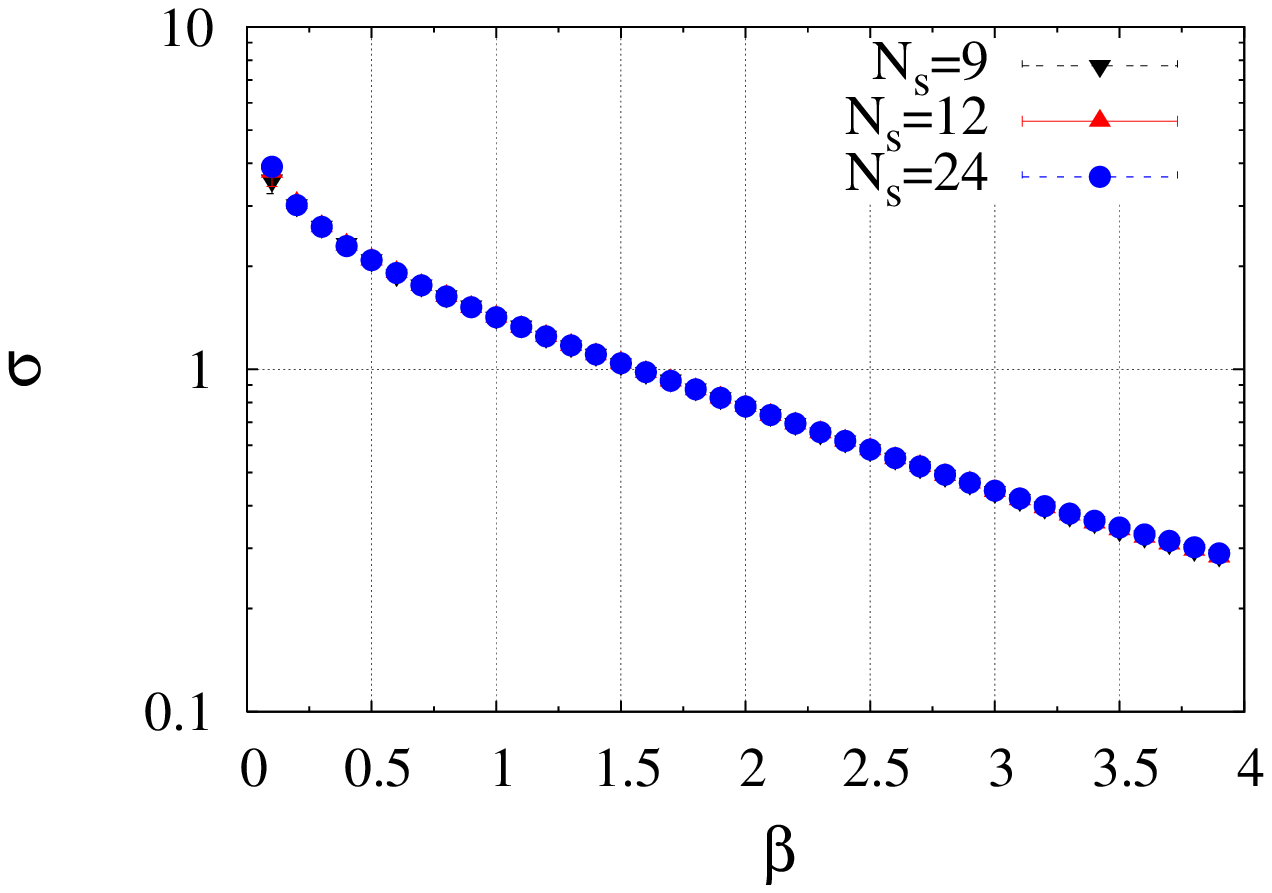}
\includegraphics*[width=0.49\linewidth]{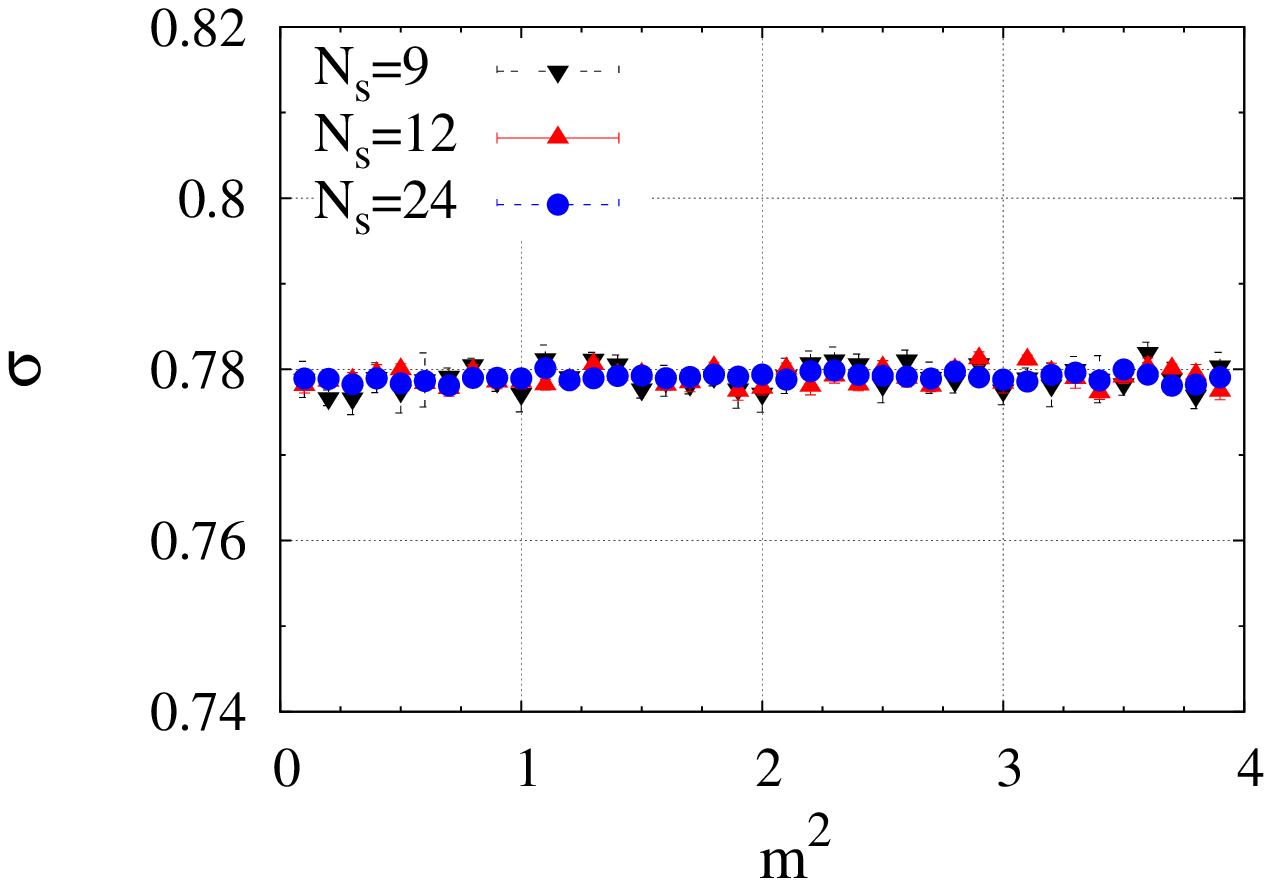}
\caption{The spatial string tension as a function of $\beta$ (left) and as a
function of $m^2$ for fixed $\beta=2.0$ (right). The string tension appears
to be independent of $m^2$. \label{fig:string_fig}}
\end{figure}

We find that both the string tension and the Wilson action drop
approximately exponentially with $\beta$ but are independent of
$m^2$. Both are also independent of the lattice size. The magnetic
sector is largely unaffected by deconfinement in the electric sector.

\end{document}